\documentclass[aps,prb,twocolumn,superscriptaddress,english]{revtex4-2}
\usepackage[utf8]{inputenc}

\usepackage{amsmath}
\usepackage{amssymb}
\usepackage{graphicx}
\usepackage[dvipsnames]{xcolor}
\usepackage{siunitx}
\usepackage{lipsum}
\usepackage{orcidlink}
\usepackage{mhchem}

\usepackage{lineno}

\newcommand{\tc}{$T_{c}$}

\newcommand{\ep}[1]{$ep$}

\newcommand{\XB}[1]{XB$_{#1}$}

\usepackage{hyperref}
\usepackage{orcidlink}
\usepackage{xr-hyper}

\newcommand*{\addFileDependency}[1]{
\typeout{(#1)}
\@addtofilelist{#1}
\IfFileExists{#1}{}{\typeout{No file #1.}}
}\makeatother

\makeatletter
\def\@fnsymbol#1{\ensuremath{\ifcase#1\or \dagger\or *\or \ddagger\or
   \mathsection\or \mathparagraph\or \|\or **\or \dagger\dagger
   \or \ddagger\ddagger \else\@ctrerr\fi}}
\makeatother

\begin{document}

\author{Simone Di Cataldo  \orcidlink{0000-0002-8902-0125}}\email{simone.dicataldo@uniroma1.it}
\affiliation{Dipartimento di Fisica, Sapienza Universit\`a di Roma, Piazzale Aldo Moro 5, 00187 Roma, Italy}
\author{Antonio Sanna \orcidlink{0000-0001-6114-9552}}
\affiliation{Max-Planck-Institut f\"ur Mikrostrukturphysik, Weinberg 2, D-06120 Halle, Germany}
\affiliation{Institut f\"ur Physik, Martin-Luther-Universit\"at Halle-Wittenberg, D-06099 Halle, Germany}
\author{Lilia Boeri \orcidlink{0000-0003-1186-2207}}
\affiliation{Dipartimento di Fisica, Sapienza Universit\`a di Roma, Piazzale Aldo Moro 5, 00187 Roma, Italy}

\title{Ambient-Pressure Superconductivity from Boron Icosahedral Superatoms}

\date{\today}
\begin{abstract}
We identify a new family of boron-rich compounds consisting of interconnected B$_{12}$ icosahedra, and electropositive guest atoms ($X$) in interstitial sites. These structures were found through first-principles crystal structure prediction at 50 GPa, where they could form, and are dynamically stable down to ambient pressure, so they could be formed under pressure, and brought back. When $X$ is a mono- or trivalent element the structures are metallic and superconducting. Predicted critical temperatures reach up to 42 K for CsB$_{12}$, rivaling MgB$_2$, the highest-$T_c$ ambient-pressure conventional superconductor. We interpret the XB$_{12}$ phase as a superatomic crystal: the B$_{12}$ units retain the icosahedral shape that they also exhibit in isolation, while forming an extended crystalline network. When X is a mono- or tri-valent atom, the system is metallic, and the B--B covalent bonding promotes strong electron-phonon coupling. Unlike MgB$_2$, where superconductivity is driven by a narrow subset of phonon modes, the XB$_{12}$ compounds exhibit broad, mode- and momentum-distributed coupling through both intra- and inter-superatomic vibrations. Our results highlight the XB$_{12}$ family as a promising platform for superconductivity and demonstrate the potential of superatoms as functional building blocks in solid-state materials design.
\end{abstract}

\maketitle

\section{Introduction}
An emerging route to discover new ambient-pressure superconductors relies on stabilizing exotic phases through synthesis at moderate pressures, followed by recovery at ambient conditions in a metastable state. Structures rich in boron and/or carbon are particularly attractive for this purpose. In fact, both elements are characterized by light atomic masses, tend to form covalent bonds, and are known to form allotropes under pressure that are recoverable at ambient conditions (e.g. diamond). Concerning superconductivity, light masses lead to high phonon frequencies, and covalent bonds to large electron-phonon matrix elements. These two are understood as they key driving factors to the 39 K critical temperature (\tc{}) of MgB$_2$ \cite{Pickett_PRL_2001_MgB2} and the record-high \tc{} of superhydrides \cite{Bernstein_PRB_2015_SH3, Heil_PRB_2015_bonding}. In addition to light masses and covalent bonds, materials must also be metallic to host superconductivity. Most covalent materials however are insulating and require doping to achieve metallicity. While some exceptions exist,  compounds become unstable even under modest doping levels \cite{Rosner_PRL_2002_LiBC, Boeri_2004_BdopedC, DiCataldo_PRB_2023_CaBH, Villa_2022_ScH3, Sanna_npj_2024_scH, Gao_2025_X4H15, Li_2025_YB2}. The problem is that electron-phonon (\ep{}) coupling, while being desirable for superconductivity, is also the main cause of structural instability if too large. This effect is particularly strong if all the coupling is concentrated on a single phonon mode. The critical question is: can we identify structural motifs that tolerate doping, maintain strong covalent bonds, and distribute coupling?

In this paper, we propose a strategy to do it based on a crystal of boron icosahedral \textit{superatoms} \cite{Khanna_1992_SAC, Jena_2018_superatoms}. A superatom is a group of atoms that coalesce in a single (meta)stable form, with properties that differ from those of the constituents. Once formed, superatoms can form stable crystals with other atoms, named \textit{superatomic crystals}. A notable example of such a system are carbon fullerenes (C$_{60}$), which form a stable \textit{fcc} lattice with metal atoms, and become superconducting at ambient pressure \cite{stephens1991structure}. 

Doping boron-icosahedral networks is a known route to strong covalent bonding and large electron-phonon coupling \cite{Mauri_PRB_2004_B12}. The key difference in our approach is that we dope at the interstitial sites rather than on the covalent bonds themselves, which has a smaller impact on the structure and preserves the B–B bond network. At the same time, the electron-phonon coupling comes from covalent intra- and inter-superatom bonds whose states sit at the Fermi energy. In the material proposed, the \ep{} coupling distributes uniformly over modes and momenta, which makes the structure able to withstand even large coupling without becoming unstable.

Through first-principles crystal structure prediction at moderate pressures (50 GPa), we found that a superatomic crystal phase \XB{12} (X is an atom of groups I--III) is stable for multiple elements. For all elements but Sc, this phase is also dynamically stable both at 50 GPa and at ambient pressure. We characterized its thermodynamic stability, electronic, and superconducting properties by means of state-of-the-art first-principles calculations. At ambient pressure, \XB{12} phases with group I and III elements exhibit superconductivity with predicted \tc{} values of up to 42 K in CsB$_{12}$.

This article is structured as follows.
In Sec.~\ref{sect:thermodynamical_stability}, we assess the thermodynamic stability of the \XB{12} family at 50~GPa, discussing their synthesizability and prospects for recovery at ambient conditions.
In Sec.~\ref{sect:superatomic_crystal}, we present the concept of superatomic crystal and discuss how \XB{12} fulfills the criteria.
In Sec.~\ref{sect:electronic}, we analyze the electronic structure \XB{12} structures in comparison with the $\alpha$-boron structure, which also consists of icosahedral superatoms. This comparison confirms that B$_{12}$ behaves consistently in different environments, as if it were a single atom, with core-like orbitals that do not interact with surrounding atoms, and valence-like ones that do.
In Sec.~\ref{sect:superconductivity_trends}, we present superconductivity trends across the full \XB{12} family, and discuss the role of electron-phonon coupling and phonon softening in determining their $T_c$.
In Sec.~\ref{sect:scdft}, we validate these results for two limiting cases, CsB$_{12}$ and LaB$_{12}$, using fully anisotropic SCDFT calculations, and analyze the Eliashberg function and the screened Coulomb interaction. Conclusions are drawn in Sec.~\ref{sect:conclusions}.

\section{Thermodynamical stability}
\label{sect:thermodynamical_stability}
We begin our discussion by looking at the thermodynamics, to show how the predicted XB$_{12}$ could form. The XB$_{12}$ structure was found performing evolutionary crystal structure prediction at 50 GPa for the XB$_{12}$ composition \cite{Oganov_JCP_2006_uspex, Lyakhov_cpc_2013_uspex} (X = K, Ca, Sc, Rb, Sr, Y, Cs, Ba, La). This pressure was chosen to access structures with exotic bonding behavior, while remaining close enough to ambient pressure to enable their recovery at ambient conditions. The specific choice of B$_{12}$ composition was motivated by the experimental report of a $Fm\bar{3}m$ phase with ZrB$_{12}$ composition, in which boron forms a continuous network \cite{Matkovich1965structure, Kennard_JSSC_1983_ZrB12}, that is superconducting at 6 K and ambient pressure \cite{Kennard_JSSC_1983_ZrB12, sluchanko2013superconductivity}. Our working hypothesis was that similar structures would appear under pressure for elements other than Zr.

For Sc and Y, we found the $Fm\bar{3}m$ structure \cite{Matkovich1965structure, Kennard_JSSC_1983_ZrB12} to be energetically favored by more than 100 meV/atom. For all other X elements investigated, we found a phase with space group $Pm\bar{3}$, composed of B$_{12}$ icosahedral units, with an enthalpy gain compared to the $Fm\bar{3}m$ phase ranging from 23 meV/atom (Ca) to 264 meV/atom (Ba). For alkaline earths, our results are consistent with a previous study, which also reported this phase for SrB$_{12}$, but did not study group I and group III elements \cite{Pu_ChinPhysLett_2021_SrB12}.
 
\begin{figure}[ht]
    \centering
	\includegraphics[width=1.0\columnwidth]{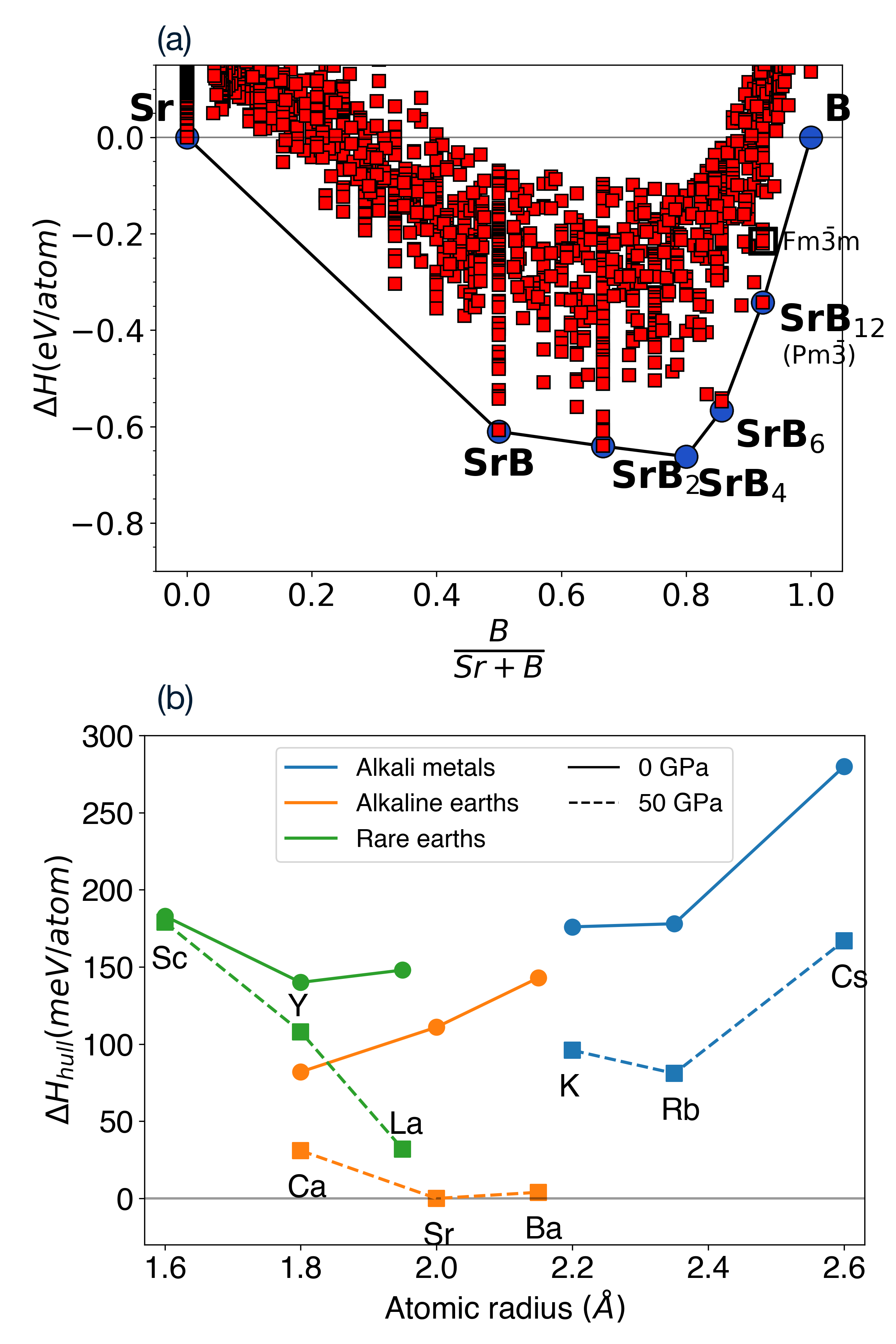}
	\caption{Thermodynamical stability of X-B boride (X = K, Ca, Sc, Rb, Sr, Y, Cs, Ba, La) phases at 50 GPa. Panel (a): convex hull of Sr-B at 50 GPa. The superatomic ($Pm\Bar{3}$) SrB$_{12}$ phase is on the convex hull. The point corresponding to the known XB$_{12}$ phase (Fm$\bar{3}$m) is indicated by a black square. Panel (b): distance from the convex hull of the superatomic XB$_{12}$ phase at 0 (solid lines) and 50 GPa (dashed lines).}
	\label{figure1}
\end{figure}

To estimate the actual synthesizability of this phase, we computed the full X-B convex hulls at 0 and 50 GPa. An example is shown for SrB$_{12}$ at 50 GPa in Fig.~\ref{figure1}(a) \cite{foot_extrahulls}: structures lying on the convex hull (blue circles) are thermodynamically stable, while those above it (red squares) are unstable, or metastable. The SrB$_{12}$ phase with $Pm\bar{3}$ space group lies on the convex hull, i.e. its enthalpy \textit{relative} to the hull ($\Delta H_{hull}$) is zero, and should form spontaneously. In the following, we will refer to the enthalpy relative to the hull as $\Delta H_{hull}$, and relative to the pure elements as $\Delta H_{f}$, respectively. The latter indicates if the structure would form from the pure elements, provided the pathway to other competing phases was inhibited.

Figure~\ref{figure1}(b) shows $\Delta H_{hull}$ across all the X elements studied, at both 0 and 50 GPa (For $\Delta H_{f}$ we refer the reader to Fig. S3 of the SM \cite{SM}). At 50 GPa, the $Pm\bar{3}$ phase lies close to the hull for four elements: Sr and Ba are stable, and Ca and La are within 50 meV/atom. At ambient pressure, the $Pm\bar{3}$ - XB$_{12}$ phase becomes less stable for all X, with $\Delta H_{hull}$ rising to 100–300 meV/atom (10-30 kJ/mol). $\Delta H_{f}$, on the other hand, at 50 GPa is negative for all elements but Rb (+10 meV/atom), and Cs (+167 meV/atom). At ambient pressure it becomes larger across the board, and becomes positive for La (+140 meV/atom), K (135 meV/atom), Rb (+198 meV/atom), and Cs (+297 meV/atom).

\begin{figure}[ht]
    \centering
	\includegraphics[width=0.95\columnwidth]{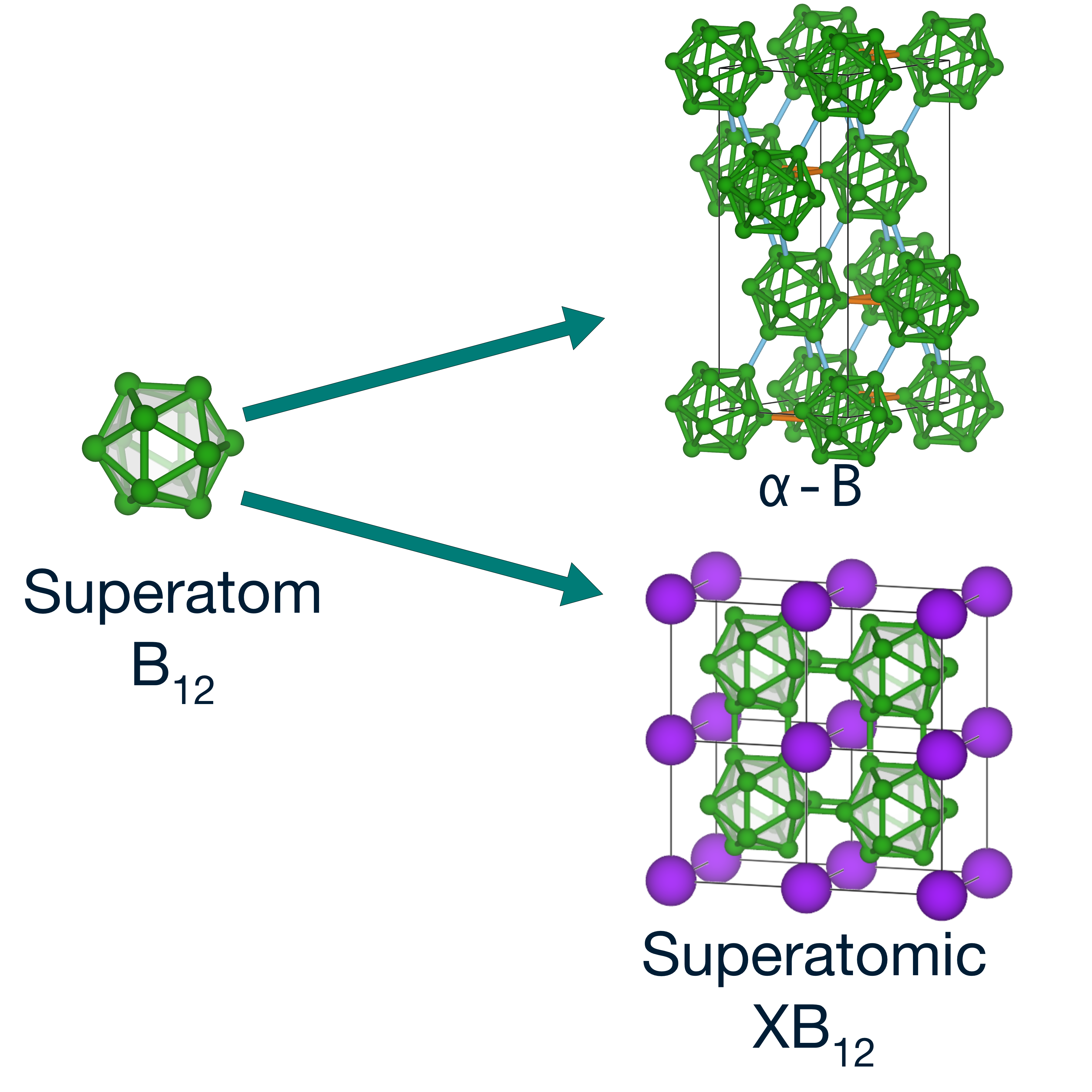}
	\caption{Formation of the XB$_{12}$ crystal from the B$_{12}$ superatomic building block. We show an isolated B$_{12}$ superatom, the crystal structure of $\alpha$-B, with the B$_{12}$ superatoms highlighted, and the $Pm\bar{3}$ - XB$_{12}$ crystal structure. In the $\alpha$-B structure, covalent and $2e3c$ bonds are highlighted in cyan and orange, respectively.}
	\label{figure2}
\end{figure}

Quantifying precisely the lifetime of \XB{12} in a metastable state would require identifying the decomposition path, which is at the moment beyond the state of the art. Here, we argue based on qualitative arguments the plausibility that the $Pm\bar{3}$ -- XB$_{12}$ phase can be synthesized under pressure and recovered at ambient conditions. First, phonon calculations confirm that most compounds (excluding Sc) are dynamically stable at zero pressure (Fig.~S11). Second, boron shares with carbon and nitrogen a propensity to form bonds with high cohesive energies. Statistically, this is associated with a window of metastability as wide as 200 meV/atom \cite{Sun_2016_SciAdv}, which is comparable to our highest $\Delta H_{hull}$ values. These factors suggest that kinetic barriers, rather than thermodynamic driving forces, are the dominant constraint on decomposition. Third, the B$_{12}$ icosahedron is a recurring motif that is present in pure boron \cite{Decker_1959_alphaB}, in boron-rich YB$_{66}$ \cite{Richards_1969_YB66}, and also as an isolated cluster \cite{Zhai_2003_B_clusters, Xu_2003_B12}. In Fig. \ref{figure2} we show the $Pm\bar{3}$ -- XB$_{12}$ structure, the B$_{12}$ icosahedron, and the ground-state structure of boron ($\alpha$-B). From a structural point of view, the \XB{12} is like a molecular crystal, with interconnected B$_{12}$ atoms, and X atoms in the interstitials. For as long as the B$_{12}$ icosahedron remains intact, the decomposition is only possible towards a phase with intact icosahedra, i.e. $\alpha$ boron. 

To support our arguments more quantitatively, we also compare our results XB$_{12}$ to those for SrB$_3$C$_3$, since the two materials present several similarities: both consist of light elements (B/C) arranged in a covalent network with large voids, where an electropositive element sits interstitially. SrB$_{3}$C$_{3}$ was experimentally synthesized under pressure and recovered at ambient conditions \cite{Strobel_SrB3C3_2017_arXiv, Strobel_AngChem_LaB3C3_2021, Wang_PRB_2021_SrB3C3_SC, Zhu_PRR_2023_B3C3, Geng_JACS_2023_B3C3}. The synthesis was achieved by laser heating a mixture of SrB$_6$, SrC$_2$, and C at a pressure of 50 GPa, where it is thermodynamically stable. After synthesis, the compound was recovered at ambient conditions simply by releasing the pressure. Since the materials are quite similar, if one was synthesized and recovered, the other one should too.

At 50 GPa SrB$_3$C$_3$ was reported on the convex hull ($\Delta H_{hull}$ = 0, $\Delta H_f$ = -0.5 eV/atom) \cite{Zhu2020_SciAdv_b3c3framework_2020}. However, at ambient pressure it exhibits a positive $\Delta H_f$ (+30 meV/atom) \cite{DiCataldo_PRB_2022_BC}. $Pm\bar{3}$ - SrB$_{12}$ is also on the convex hull at 50 GPa, and at ambient pressure exhibits a negative formation enthalpy ($\Delta H_f$ = -110 meV/atom). In comparison, SrB$_{12}$ is more stable. The synthesis pressure is about the same and, since the formation energy is negative, should be equally recoverable at ambient conditions. 

We also propose a possible synthesis route of XB$_{12}$ which does not require pressure. In fact, $\alpha$-B already contains preformed B$_{12}$ icosahedra. In principle, one could mix a finely-ground powder of $\alpha$-B and the selected X element, and apply only moderate heat to promote diffusion of X atoms among B$_{12}$ icosahedra, while leaving those intact. If the B$_{12}$ superatoms do not break, the reaction toward phases with a lower boron content is inhibited, and XB$_{12}$ can form as long as $\Delta H_{f}$ is negative, since no other phase we found contains B$_{12}$ icosahedra.

\section{XB$_{12}$ as superatomic crystals}
\label{sect:superatomic_crystal}

\begin{figure*}[ht]
    \centering
	\includegraphics[width=1.75\columnwidth]{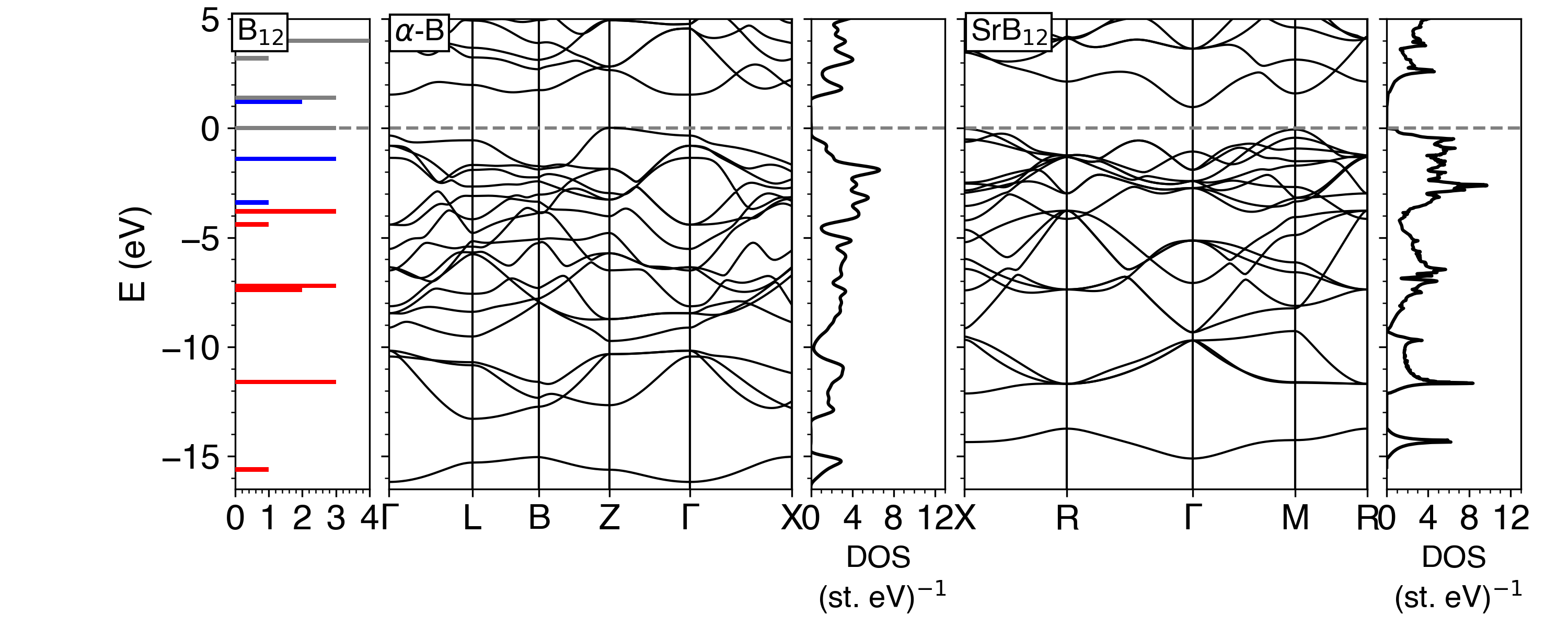}
	\caption{Calculated electronic structure and total density of states (DOS) for an isolated B$_{12}$ superatom, $\alpha$-B, and SrB$_{12}$ at ambient pressure. The energy zero is set at the top of the valence band. In the DOS panel for B$_{12}$ we indicate with red the B$_{12}$ molecular states, and with blue and gray the nonbonding states that in the XB$_{12}$ crystal become occupied and unoccupied, respectively.}
	\label{figure3} 
\end{figure*}
We now discuss the crystal structure of the $Pm\bar{3}$ - \XB{12} phase, show how the B$_{12}$ icosahedron appears as a rigid building block, and discuss how it matches the definition of a superatom. This idea will be key to later understand the materials' electronic and vibrational properties.

A superatom is defined as an assembly of atoms that exhibits one or more atom-like properties \cite{Jena_2018_superatoms}. This definition is intentionally broad and encompasses small molecules, radicals, and clusters. Assembling superatoms into periodically ordered \textit{superatomic crystals}, where superatoms retain their identity while exhibiting collective behavior, is a major challenge in physical chemistry \cite{Khanna_1992_SAC, Khanna_1995_SAC2}. 

Superatomic crystals are characterized by two distinct length scales: an intra-superatomic scale (the size of each superatom), and an inter-superatomic one (the distance between superatoms). This translates into two features that determine the electronic structure: discrete energy levels coming from orbitals localized on individual superatoms, and inter-superatom hopping. The energy levels corresponding to localized states are low-lying, and analogous to atomic semi-core states. Those that arise from inter-superatom hopping instead lead to dispersive bands, analogously to the inter-atom hopping of regular crystal. Depending on the inter-superatom distance, the hopping can be as large as for regular atoms, even though it involves molecular orbitals. 

In the XB$_{12}$ structure, B$_{12}$ superatoms form twelve inter-superatom (B--B) bonds with their outward-facing orbitals. The inter- and intra-superatom distances are both around 1.7 \AA{}, hence the inter-superatom electronic hopping is comparable to the intra-superatom one. 

To support the idea that the \XB{12} structure does qualify as a superatomic crystal, we compare it with the ground-state structure of boron ($\alpha$-B). There, B$_{12}$ icosahedra are connected with their neighbors along the $z$ axis through six B--B bonds, also 1.7 \AA{} long, and to their neighbors in the $xy$ plane via six two-electrons, three-centers bonds ($2e3c$) \cite{Albert_2009_boron, Zhou_2014_unexpected}. So in the two structures the B$_{12}$ superatoms maintain well-defined geometries and bond lengths, while the periodic crystal is formed through inter-superatom bonds, which differ in coordination and geometry. 

Concerning our specific goal of identifying conventional superconductors, the XB$_{12}$ superatomic crystal possesses two important features. The first is that both inter- and intra-superatom bonds involve the same atoms (B), hence must be covalent \cite{Pickett_PRL_2001_MgB2}. The second feature is the ability of \XB{12} to withstand electron or hole doping over a broad range without becoming dynamically unstable. This aspect is discussed in more detail in Sect. \ref{sect:superconductivity_trends}; the phonon dispersions at ambient pressure for all structures considered are shown in Fig. S11 of the SM \cite{SM}.

We ascribe the stability of the \XB{12} crystal to two geometric factors. First, doping occurs through substitution at the interstitial X site, which in the system acts as a charge donor. This should be easier to realize experimentally, compared to doping the covalent bond (e.g. by C substitution) \cite{Mauri_PRB_2004_B12, Flores_2018_poly, DiCataldo_PRB_2023_CaBH}. Second, the B$_{12}$ superatoms present \textit{internal} degrees of freedom. Therefore, the B$_{12}$ can respond to doping with slight distortions, while the overall crystal structure remains unchanged, something that is not possible in regular crystals. 

\section{Electronic structure}
\label{sect:electronic}
We now describe the electronic structure of \XB{12} in comparison with $\alpha$-B, and the isolated B$_{12}$ superatom. This comparison reveals that the low-lying electronic states of the B$_{12}$ do not change when their arrangement changes, and supports the idea that the electronic behavior of the B$_{12}$ icosahedron is atom-like.

To better understand the electronic structure of B$_{12}$, let us begin by considering it in isolation. Its molecular orbitals can be obtained from the 2$s$ and 2$p$ orbitals of the 12 boron atoms. These result in a total of 48 molecular orbitals, of which 13 bonding, with $A_{g}$, $T_{1u}$, $H_{g}$ symmetries (1, 3, and 9 molecular orbitals, respectively), 12 nonbonding, and 13 anti-bonding \cite{Longuet_1955_B12_MO}. These orbitals must then be filled with 36 electrons (3 per boron). 
The remaining 10 electrons fill the non-bonding orbitals ar higher energies. These orbitals face outward \cite{Longuet_1955_B12_MO}, and are those that form inter-superatomic bonds before they evolve into bands. These orbitals lead to different bonds in \XB{12} and $\alpha$-B, in a way that depends on the geometry.

In Fig.~\ref{figure3} we show the electronic structure for the isolated B$_{12}$, compared with SrB$_{12}$ and $\alpha$-B at ambient pressure, along with the total density of states (DOS). In the isolated B$_{12}$, one can count the 13 bonding states, going from -15.5 to -3.9 eV, followed by partially-filled non-bonding states. 

In $\alpha$-B there are 18 valence bands. It is immediately visible that the first, isolated band at -15 eV, and the group of three bands around -10 eV match those of the isolated atom both in energy and electron count. According to Ref. \cite{Zhou_2014_unexpected}, nine of the remaining bands derive from the other bonding states, and the last five from 3 covalent bonds and 2 two-electrons-three-centers bonds. The 36 electrons of boron completely fill these orbitals, hence the system is a semiconductor.

For the $Pm\bar{3}$-XB$_{12}$ compounds, the essential features of the electronic structure do not depend on X \cite{footsemicore}. For all X elements the band structure follows a rigid band behavior (For all the bands we refer the reader to Fig. S7 of the SM \cite{SM}). When X is an alkali metal (group III element) the system is hole-doped in the valence band (electron-doped in the conduction band). In XB$_{12}$, the valence bands are 19. The single band at -15 eV, and the group of three around -10 are immediately recognizable, like in $\alpha$-B. Focusing on the $\Gamma$ point, one can count a total of nine bands in the range from -9.5 to -3.5 eV, which correspond to the remaining B$_{12}$ bonding orbitals (For the full evolution of the eigenvalues from isolated B$_{12}$ to SrB$_{12}$ we refer the reader to Fig. S4 of the Supplemental Material \cite{SM}). The remaining 6 bands arise from the 12 inter-superatom covalent bonds. The 36 electrons from B fill 18 bands, leaving one unfilled. As a result,  when X is a divalent atom the system is insulating. It is also the valence of the X atom for which the system is thermodynamically most stable. The DFT band gap at ambient pressure is 1 eV for Ca, 1.5 eV for Sr, and 1.7 eV for Ba. A mono/trivalent X gives the metallic state needed for superconductivity.

In other words, we showed that the bonding molecular orbitals of B$_{12}$ behave similarly to semi-core atomic states, which are only slightly perturbed by the environment, while the non-bonding ones behave analogously to valence ones, which change their behavior depending on the environment. 

We finish this section by discussing how the bonds that are involved in superconductivity appear in real space, which revealed that both inter- and intra-superatomic bonds participate.

\begin{figure}[ht]
    \centering
	\includegraphics[width=0.95\columnwidth]{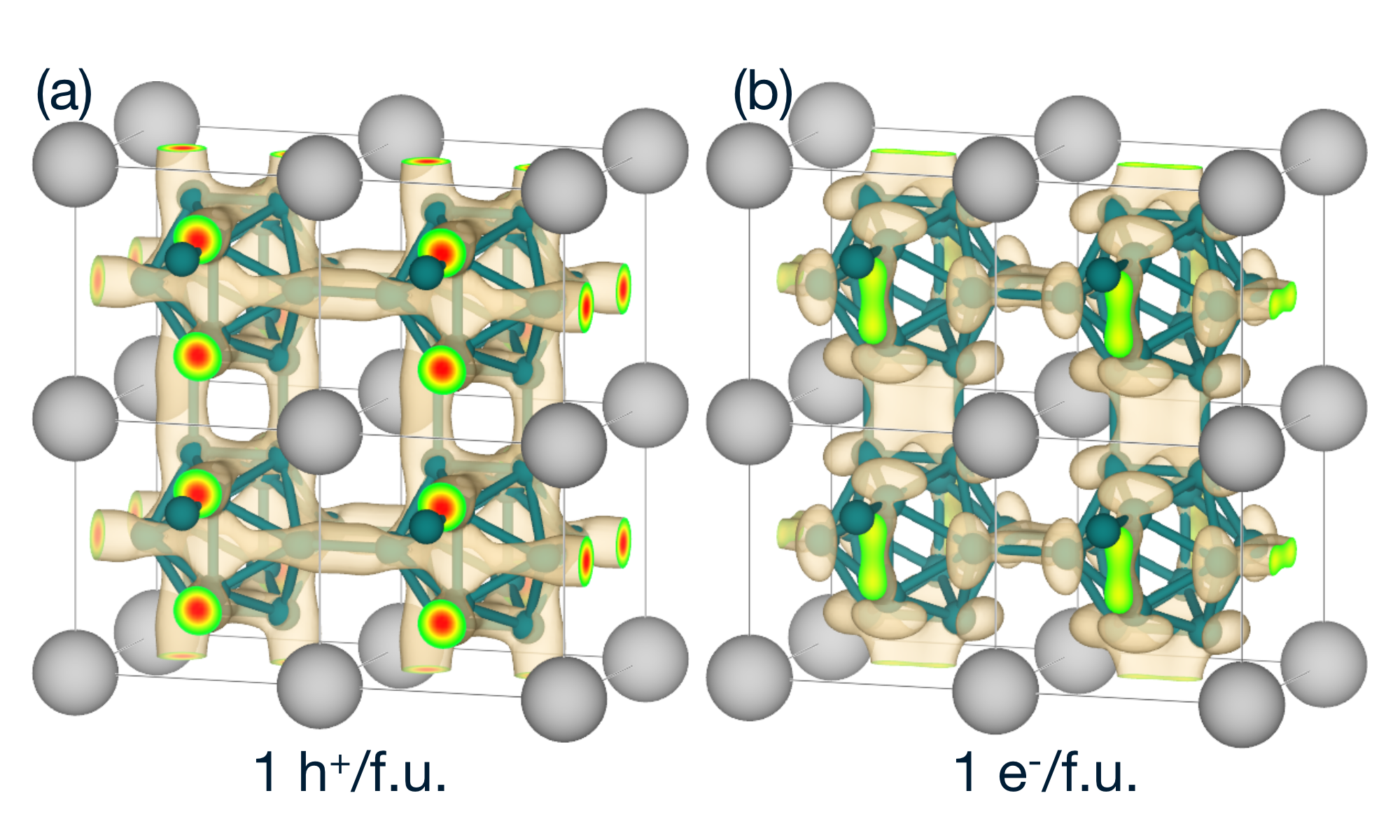}
	\caption{Panel (a): local DOS for SrB$_{12}$ with one extra hole per formula unit. Panel (b): local DOS for SrB$_{12}$ with one extra electron per formula unit.}
	\label{figure4} 
\end{figure}

The only electronic states that contribute to electron-phonon coupling are those near the Fermi level. In Fig.~\ref{figure4} (a) and (b) we show the local DOS (the real-space localization of the DOS) at the Fermi level under hole and electron doping, respectively (for a detailed definition we refer the reader to Sect. S1 C of the SM \cite{SM}). In the hole-doped case, states localize over both the inter- and intra-B$_{12}$ bonds. In the electron-doped case, states at the Fermi energy localize primarily on the single boron atoms, and in the square formed by two inter-B$_{12}$ bonds. In both cases, electron-phonon coupling involves contributions from both intra- and inter-superatom bonds, which indicates that the all the degrees of freedom of superatoms contribute to superconductivity.

In both the electron- and hole-doped case, the states near the Fermi level -- just above and just below the gap, respectively -- are of predominantly boron character (See Fig. S5 and S6 of the SM for the atom-projected DOS for all X atoms \cite{SM}). These localized orbitals give rise to a strong electron–phonon coupling, comparable to that of MgB$_2$. In fact, the concept of covalent-bonds driven metallic, originally proposed for MgB$_2$~\cite{Pickett_PRL_2001_MgB2}, applies equally well to these materials. We will quantify this in the next section. 

\begin{figure}[ht]
    \centering
	\includegraphics[width=1.0\columnwidth]{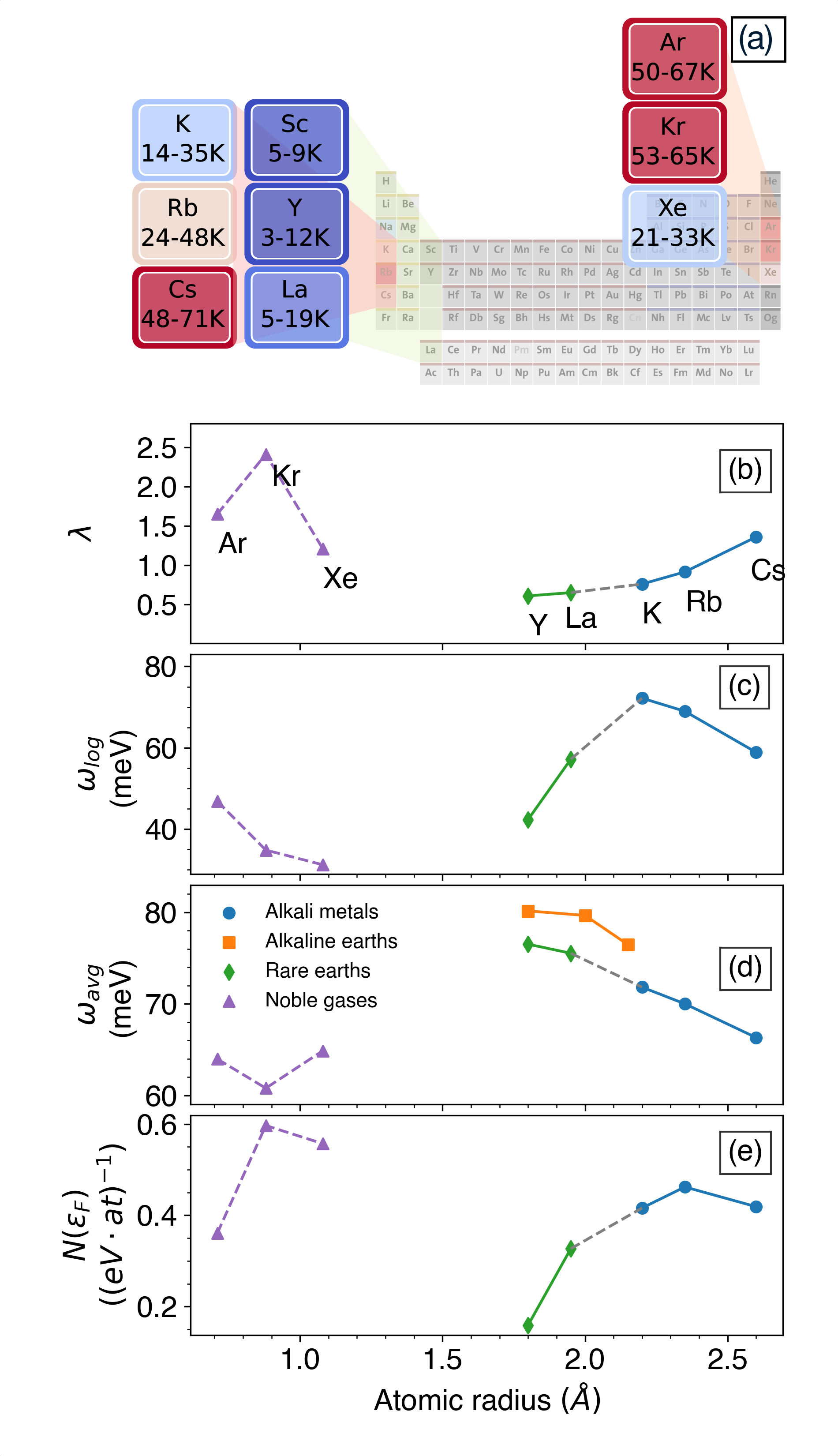}
	\caption{Main superconducting observables for metallic XB$_{12}$ as a function of the radius of the guest atom. Panel (a) shows the McMillan \tc{} (in a range of $\mu^{*}$ between 0.1 and 0.2) with respect to the X element in the periodic table. Panels (b), (c), (d), and (e) show the electron-phonon coupling $\lambda$, the logarithmic average of the phonon frequency $\omega_{log}$, the plain average of the phonon frequency $\omega_{avg}$, and the DOS at the Fermi energy $N(\epsilon_F)$, respectively. Alkali metals, alkaline earths, rare earths, and noble gases are shown as blue circles, orange squares, green diamonds, and purple triangles, respectively.}
	\label{figure5}
\end{figure}

\section{Superconductivity trends}
\label{sect:superconductivity_trends}
We now move on to discuss the superconducting properties of the $Pm\bar{3}$ - \XB{12} family. In this section we focus on the \tc{} trends across the family, to understand how much it is possible to optimize it. For this reason, we use the simple but transparent McMillan formula to estimate the \tc{}. In Sect. \ref{sect:scdft}, we will focus on the two most promising compounds -- LaB$_{12}$ and CsB$_{12}$ using state-of-the-art Density Functional Theory for Superconductors. 

The McMillan formula is \cite{McMillanTC, Allen_PRB_1975_transition}:
\begin{equation}
\label{eq:mcmillan}
T_c = \frac{\omega_{\log}}{1.2} \exp\left[
- \frac{1.04 (1 + \lambda)}{\lambda - \mu^*(1 + 0.62\lambda)}
\right];
\end{equation}
here $\omega_{\log}$ is the logarithmic average phonon frequency, $\lambda$ is the electron-phonon coupling constant, and $\mu^*$ is the Coulomb pseudopotential. According to this expression, \tc{} increases with strongly with $\lambda$ in the 0-2 range, increases linearly with $\omega_{log}$, and decreases with $\mu^{*}$. In general, $\lambda$ is the dominating factor, but an increase in coupling is always associated with a suppression of the characteristic phonon energy $\omega_{\log}$. Beyond a certain threshold, the reduction in $\omega_{\log}$ offsets the gain from increased $\lambda$, and can even lead to a structural instability - a behavior characteristic of the Cohen–Anderson limit~\cite{Cohen_AIP_CP_1972_limit_Tc}.

An intuitive description of $\lambda$ is given by Hopfield's approximated expression~\cite{Hopfield_PhysRev_1969_superconductivity}:
\begin{equation}
\label{eq:hopfield}
\lambda = \frac{N(\epsilon_F) \langle g^2 \rangle}{M \langle \omega^2 \rangle}.
\end{equation}
Where $N(\epsilon_F)$ $\langle g^2 \rangle$,  $M$ and $\langle \omega^2 \rangle $ are the DOS at the Fermi energy, electron-phonon matrix element, atomic mass and average phonon frequency, respectively. $M$ and $\langle\omega^2 \rangle$ depend on the material, while $\langle g^2 \rangle$ mainly depends on which bonds contribute to states at the Fermi energy, and $N(\epsilon_F)$ depends on the position of the Fermi energy.

Fig. \ref{figure5} summarizes the main superconducting properties for all metallic XB$_{12}$ compounds at ambient pressure. Critical temperatures were predicted using Eq. \ref{eq:mcmillan} and a $\mu^{*}$ of 0.10 and 0.20, which is a conservative range of values assumed by most materials \cite{Pellegrini_SimplifiedEliashberg2022, Pellegrini_NatRev2024}. The calculated \tc{}'s range from 10 K for trivalent X (Y, La) to over 70 K for CsB$_{12}$. 

Using Eq. \ref{eq:mcmillan}, it is clear that this variation is caused by the corresponding increase in $\lambda$ (panel (b)), from $\sim 0.7$ in LaB$_{12}$ to $\lambda \sim 1.5$ (strong coupling) in CsB$_{12}$. As shown in Fig.~\ref{figure5} (e), the trend in $\lambda$ closely follows the behavior of the electronic density of states (DOS) at the Fermi level $N(\epsilon_F)$, hence if we could increase it, $\lambda$ should follow, and enhance the \tc{}.

We discussed in Sect. \ref{sect:electronic}, that the whole \XB{12} family follows a rigid band behavior (Fig. S7 of the SM \cite{SM}). Given the steep profile of the DOS, one could envision further boosting the DOS via hole doping. However, introducing Cs vacancies into CsB$_{12}$ would leave a large void in the structure, which would lead to its collapse under pressure. Instead, to test the idea of hole doping in a more controlled way, we performed a theoretical study of hypothetical XB$_{12}$ compounds with noble gas elements (Ar, Kr, Xe) at the X site.

Although these compounds exhibit prohibitively high formation enthalpies ($ > 500$~meV/atom) and are not synthesizable, they provide a useful test of the electronic mechanism to boost \tc{}. Indeed, noble gas substitutions significantly increase both $N(0)$ and $\lambda$. Yet, despite these gains, the predicted \tc{} remains essentially unchanged. This is due to the concurrent phonon softening, which offsets the enhanced coupling strength. The phonon softening is visible in the drop of the average phonon frequency $\omega_{avg}$ (panel (d)), that accompanies the increase in $\lambda$ (panel (b)). Unlike $\omega_{log}$, where modes with more coupling weigh more \cite{Allen_PRB_1975_transition}, $\omega_{avg}$ weighs each mode equally. Comparing $\omega_{avg}$ for alkaline-earth members (insulating, orange symbols), for which there is no coupling, and the rest (metallic, green, blue and purple symbols), reveals that coupling alone shifts the average frequency by 20 meV. In principle, phonon softening could also originate from hole doping that empties bonding states, effectively weakening the bond stiffness and hence decreasing the (bare) phonon frequencies. However, the strong dependence of $\lambda$ observed in Fig. \ref{figure5} (b) within atoms of the same group (Ar-Kr-Xe; K-Rb-Cs), where the charge remains constant, indicates that the coupling is the dominating factor.

Because the decrease in $\omega_{log}$ compensates the increase in $\lambda$, the \tc{} of CsB$_{12}$ is already at the saturation limit for the \XB{12} family. As we will see in the following, the presence of a gap right above the Fermi energy has also consequences on $\mu^{*}$, which is anomalously large, due to reduced screening \cite{Morel_PhysRev_1962_mustar,SSW_StrongCouplingSC_PR1966}. As a result, despite strong coupling and favorable DOS, the critical temperatures of these new borides remain moderate. In this sense, CsB$_{12}$ lies near the structural, dynamical, and electronic optimum of the XB$_{12}$ family.


\section{Superconducting properties of \ce{CsB12} and \ce{LaB12} from Density Functional Theory for Superconductors}
\label{sect:scdft}
In this section we focus on the two compounds that appeared as the most promising in the previous section: CsB$_{12}$ (monovalent atom, highest \tc{}), and LaB$_{12}$ (trivalent atom, most likely to be synthesized, lowest \tc{}). We examine their vibrational and superconducting properties in detail, identify the factors that enable \XB{12} to support superconductivity, and validate the results of the previous section.

To study CsB$_{12}$ and LaB$_{12}$ with a parameter-free theory, we solved the fully anisotropic equations of SCDFT (DFT for superconductors). SCDFT permits us to obtain the superconducting critical temperature and gap, accounting on equal footing for phonon-mediated pairing, Coulomb repulsion, and gap anisotropy effects~\cite{Pellegrini_NatRev2024}. 

\begin{figure}[ht]
    \centering
	\includegraphics[width=1.0\columnwidth]{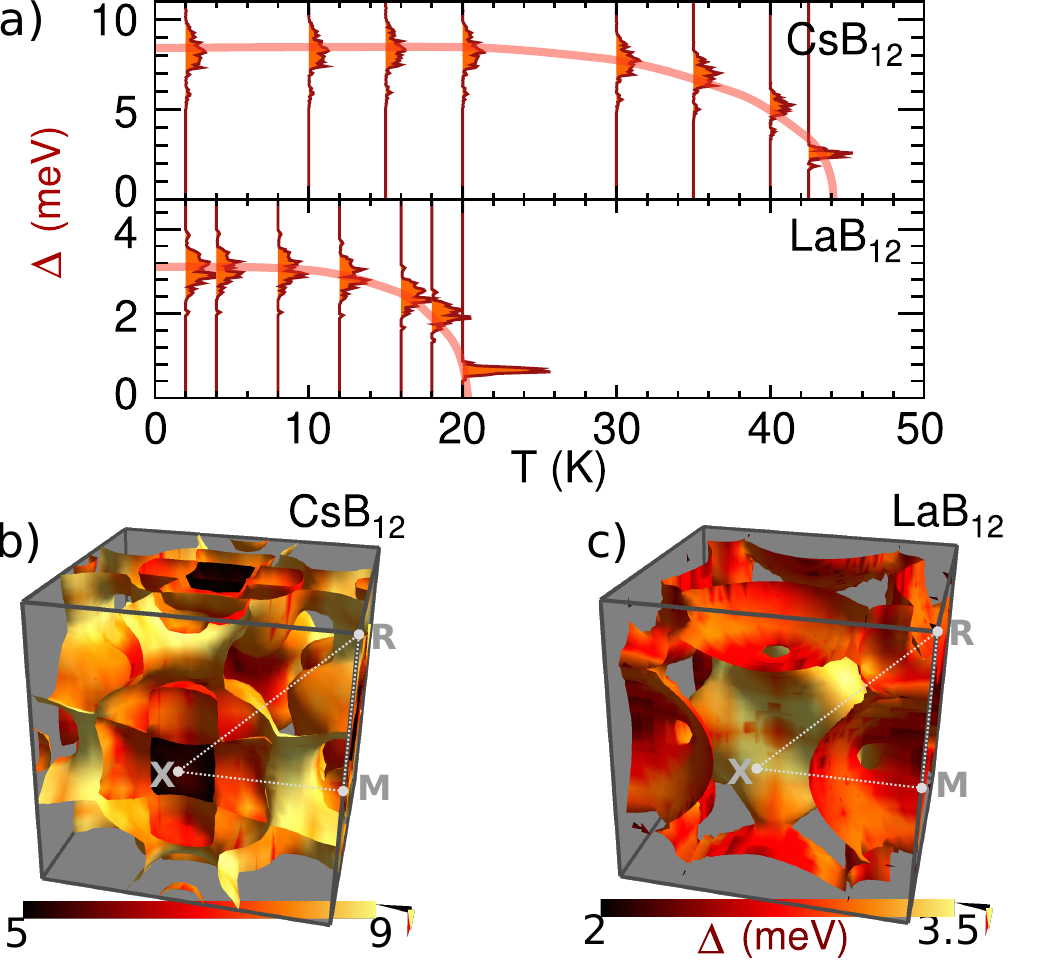}
    \caption{Superconducting gap of CsB$_\text{12}$ and LaB$_\text{12}$. a) Histograms of the anisotropic superconducting gaps as a function of temperature. The thick orange line is a guide to the eye.
    b) Fermi surfaces with, color coded, superconducting gap (at T=2K).}
	\label{figure6}
\end{figure}

Fig.~\ref{figure6} displays the temperature-dependent superconducting gap (a) and its distribution over the Fermi surface (b). In both compounds, the superconducting gap is nearly isotropic, suggesting that the pairing interaction is uniformly distributed in momentum space, without strong band- or mode-selective enhancements. The calculated critical temperatures are 42~K for CsB$_{12}$ and 21~K for LaB$_{12}$ -- values that lie at the lower and upper limit of the range predicted with the McMillan formula for the two compounds, respectively. This result suggests that Coulomb interactions in both these systems are strong, while a ratio $\frac{2\Delta}{T_c}$ is 4.2 and 3.7 suggests that coupling is on the moderate/weak regime.

\begin{figure}[ht]
    \centering
	\includegraphics[width=0.9\columnwidth]{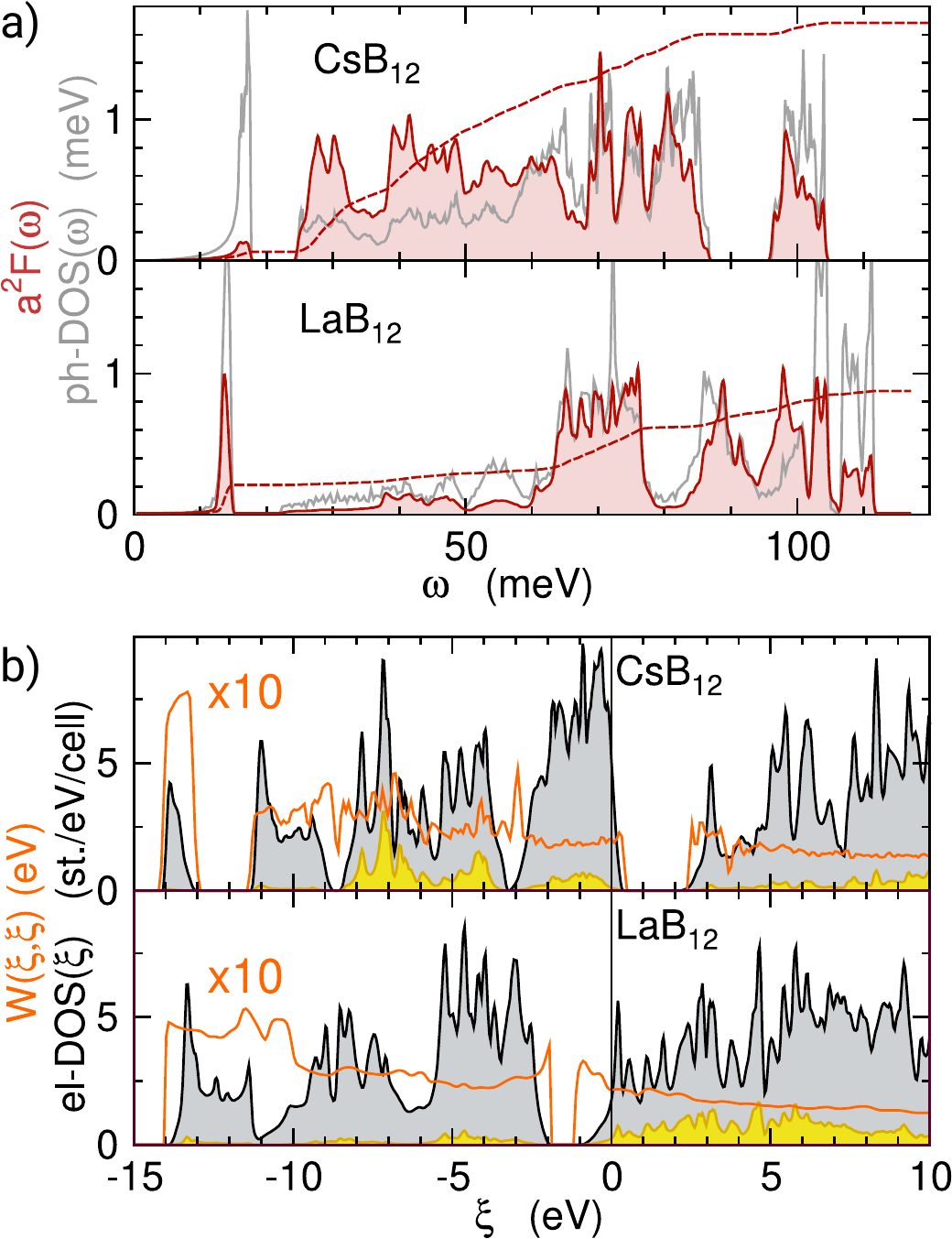}
    \caption{Electron-phonon and electron-electron (Coulomb) interaction. a) $\alpha^2F$ function of CsB$_{12}$ and  LaB$_{12}$ at 0 GPa. Phonon density of states are shown in gray and electron-phonon spectral function ($\alpha^2F$) is shown in red. The red dashed line is the integration curve of $2\alpha^2F/\omega$ leading to the coupling parameter $\lambda$.
    b) Density of electronic states (black curve) and its projection on Cs and La (yellow). The orange curve is the diagonal part $W(\xi,\xi)$ the RPA screened electronic interaction entering the SCDFT gap equation~\cite{Pellegrini_NatRev2024}.
    }
	\label{figure7}
\end{figure}

We move to analyze the two components of the pairing kernel: the screened electron–electron repulsion, and the attractive electron–phonon interaction, described by the Eliashberg spectral function $\alpha^2F(\omega)$\ and the static Coulomb interaction $W(\xi,\xi')$ ~\cite{Pellegrini_NatRev2024,AllenMitrovic1983,Flores_PR_2020perspective}, shown in Fig. \ref{figure7} (a) and (b), respectively.

The effect of Coulomb interaction is anomalously larger than most materials \cite{PellegriniKukkonenSanna_beyondRPA_PRB2023, Pellegrini_NatRev2024}. In fact, the partial gapping of states near the Fermi level reduces the efficiency of dynamical Coulomb renormalization ~\cite{Morel_PR_1962_pseudopotential,SSW_StrongCouplingSC_PR1966}. This leads to an enhancement of the effective Coulomb interaction relative to constant-DOS models.
In fact, the SCDFT critical temperatures fall at the lower end of the McMillan–Allen–Dynes predictions — indicating a high effective $\mu^*$.

Nonetheless, the screened Coulomb interaction $W(\xi,\xi')$, shown in Fig.\ref{figure7} (b), displays the smooth, featureless profile characteristic of good metals\cite{Eliashberg_GenuineJPSJ2018, PellegriniKukkonenSanna_beyondRPA_PRB2023}.
The strength of Coulomb repulsion can be quantified as $\mu = W \cdot \mathrm{DOS}$ at the Fermi level, yielding values of 0.55 for CsB$_{12}$ and 0.46 for LaB$_{12}$. While relatively large, these values are consistent with the high DOS.

The Eliashberg function $\alpha^2F(\omega)$ shown in Fig. \ref{figure7} (a) is broadly distributed over the phonon spectrum, both in terms of mode and momentum. As a result, the integrated coupling $$\lambda(\omega) = 2\int_{0}^{\omega} \frac{\alpha^2F(\omega')}{\omega'}d\omega'$$ (with $\lambda(\infty) = \lambda$)
shown as a red line in Fig. \ref{figure7} increases rather linearly with $\omega$, up to 1.7 in CsB$_{12}$ and 0.9 in LaB$_{12}$. In LaB$_{12}$, the phonon modes are generally harder than in CsB$_{12}$, as visible from the larger $\omega_{avg}$ in Fig. \ref{figure5} (d). Even though the total coupling $\lambda$ for LaB$_{12}$ is less, $\omega_{\log}$ is smaller. This results from a different distribution of the Eliashberg function, as in LaB$_{12}$ about 25\% of the total coupling comes from phonon modes at 15 meV (for the complete phonon dispersions with electron-phonon coupling we refer the reader to Fig. S10 of the Supplementary Material).

In other superconductors, such as MgB$_{2}$, the electron-phonon coupling is often strongly mode- and momentum-selective, and concentrated in a narrow frequency range. Since a single mode carries all spectral weight, it is strongly renormalized. Even if one could increase coupling, the system would rapidly become unstable and undergo a structural transition. In contrast, in the XB$_{12}$ structure this effect is mitigated by the interconnected superatoms, whose phonons involve simultaneous deformations of both intra- and inter-superatom bonds along multiple directions. As a result, the electron-phonon interaction is spread more uniformly across phonon branches and throughout the Brillouin zone, reflecting the absence of a privileged direction and distance for the interaction.

\section{Conclusions}
\label{sect:conclusions}
In conclusion, in this work we identify a class of promising boron-rich superconductors: $Pm\bar{3}$ - XB$_{12}$ compounds consisting of B$_{12}$ icosahedral superatoms and interstitial electropositive elements, assembled in a \textit{superatomic crystal}. These compounds can be synthesized under pressures of 50 GPa and, based on dynamical stability, could be recoverable at ambient conditions. \XB{12} crystals present unique advantages for superconductivity: first, boron atoms bond covalently both intra- and inter-superatom, leading to large electron-phonon matrix elements. Second, they are able to withstand extreme hole- and electron doping, which allows a fine tuning of the Fermi energy without compromising the structural stability. In fact, our DFT and SCDFT calculations predict superconducting critical temperatures up to 42 K at ambient pressure -- rivaling MgB$_2$ -- with electron-phonon coupling distributed broadly in mode and momentum space.

An alternative nonequilibrium synthesis route may be also be possible at ambient conditions. In fact, the B$_{12}$ superatoms are already present in pure $\alpha$-B, and may be directly broken apart, and intercalated with metal atoms. Compared to MgB$_{2}$, \XB{12} presents the advantage of a more isotropic structure, as well as an isotropic superconducting gap, which would represent two important advantages for applications in devices.

The potential to exploit the coupling of molecular states in a framework where the Fermi energy can be varied easily has been proposed as a promising route to find superconductors \cite{Moussa_PRB_2008_fragments}. These findings highlight the potential of superatomic crystals to realize this idea. 

\section{Methods}
Structural searches were performed using the variable-composition evolutionary algorithm implemented in the USPEX code \cite{Oganov_JCP_2006_uspex, Lyakhov_cpc_2013_uspex}, using five steps for structural relaxations. The DFT calculations were performed using the Vienna Ab-initio Simulation Package (VASP) \cite{Kresse_PRB_1993_vasp1, Kress_PRB_1996_vasp2, Kresse_PRB_1999_vasp3}, with the projector-augmented wave pseudopotentials provided with the code, and the Perdew-Burke-Ernzerhof approximation \cite{perdew_generalized_1996}. Further details are provided in the Supplementary Material \cite{SM}.

Electronic structure, phonon and electron-phonon calculations were performed using Quantum ESPRESSO \cite{Giannozzi_JPCM_2009_qe, Giannozzi_JPCM_2017_qe}. We employed Optimized norm-conserving Vanderbilt pseudopotentials \cite{Hamann_PRB_2013_ONCV}. For the ground-state charge density, we employed a cutoff of 80 Ry on the plane wave expansion, and a 8$\times$8$\times$8 Monkhorst-Pack mesh and a smearing of 0.04 Ry for integrals over the Brillouin zone. Densities of states were computed non-self-consistently using the tetrahedron method over a 24$\times$24$\times$24 grid for integrals over the Brillouin zone. 

Phonon calculations for LaB$_{12}$ and CsB$_{12}$ were performed on a 6$\times$6$\times$6 grid, and properties were also interpolated on a 18$\times$18$\times$18 grid for both electrons and phonons. Convergence tests (See an example in Fig. S14 of the SM \cite{SM}) show that integration of phonon dispersions is already reasonably converged with 2$\times$2$\times$2 grids, as a consequence of the uniformity of coupling in \textbf{q}.

The Coulomb interaction for the SCDFT has been evaluated within the random phase approximation with an energy cutoff for the band summation of about 50 eV above the Fermi level and a maximum $|{\bf G}|=3.0$ a.u. for the summation of crystal field factors~\cite{ElkCode}.  

Anisotropic simulations are done with a Monte-Carlo algorithm using 50 K k-points accumulated with higher probability near the Fermi surface~\cite{Sanna_NbSe2_npjQM2022}. The same algorithm is used to compute the electron-phonon spectral functions shown in Fig.~\ref{figure7}). Electronic and coupling parameters are linearly interpolated on the random-mesh which is used for the solution of the SCDFT Kohn-Sham gap equation. We adopt the functional from  Ref.~\cite{Sanna_PRL_2020_combining}, also used for the calculation of the physical superconducting gaps~\cite{Pellegrini_NatRev2024}. Eliashberg simulations including Coulomb interactions are performed using the approach of Ref.~\cite{Pellegrini_SimplifiedEliashberg2022}. 

Fermi surfaces are plotted with the Fermisurfer code~\cite{Kawamura_Fermisurfer19}.

\section*{Acknowledgements}
L.B. and S.D.C. acknowledge computational resources from the EuroHPC project "EXCHESS" (EHPC-REG-2024R01-089) and funding from the European Union - NextGenerationEU under the Italian Ministry of University and Research (MUR), “Network 4 Energy Sustainable Transition - NEST” project (MIUR project code PE000021, Concession Degree No. 1561 of October 11, 2022) - CUP B53C22004070006.
\bibliography{library}
\end{document}